\documentclass[11pt]{article}

\usepackage{graphicx,epsfig}
\usepackage{multicol}
\usepackage{amsmath,amssymb,cite,color,hyperref}
\newcommand{\be}{\begin{equation}}
\newcommand{\ee}{\end{equation}}
\newcommand{\bea}{\begin{eqnarray}}
\newcommand{\eea}{\end{eqnarray}}
\newcommand{\bi}{\begin{itemize}}
\newcommand{\ei}{\end{itemize}}
\newcommand{\ben}{\begin{enumerate}}
\newcommand{\een}{\end{enumerate}}

\newcommand{\lc}{\left[}
\newcommand{\rc}{\right]}
\newcommand{\lp}{\left(}
\newcommand{\rp}{\right)}

\def\frac#1#2{{{#1}\over {#2}}}
\def\gsim{\mathrel{\rlap{\lower4pt\hbox{\hskip1pt$\sim$}}
    \raise1pt\hbox{$>$}}}         
\def\lsim{\mathrel{\rlap{\lower4pt\hbox{\hskip1pt$\sim$}}
    \raise1pt\hbox{$<$}}}         

\newcommand{\rep}{\mathrm{rep}}

\newcommand{\draft}[1]{}

\definecolor{grey}{rgb}{0.5,0.5,0.5}

\begin{document}
\begin{flushright}
IFUM-974-FT\\
Edinburgh 2011/11\\
FR-PHENO-2011-005\\
TTK-11-07\\
\end{flushright}
\begin{center}
{\large\bf Precision determination of
 $\alpha_s\lp M_Z\rp$  \\using an unbiased global
NLO parton set}
\vspace{0.8cm}

{\bf  The NNPDF Collaboration:}\\
Simone~Lionetti$^1$, Richard~D.~Ball$^{3}$, Valerio~Bertone$^4$, Francesco~Cerutti$^5$, \\
 Luigi~Del~Debbio$^3$, Stefano~Forte$^{1,2}$, Alberto~Guffanti$^4$, 
Jos\'e~I.~Latorre$^5$, \\ Juan~Rojo$^{1,2}$ and Maria~Ubiali$^{6}$.

\vspace{1.cm}
{\it ~$^1$ Dipartimento di Fisica, Universit\`a di Milano and\\
~$^2$ INFN, Sezione di Milano,\\ Via Celoria 16, I-20133 Milano, Italy\\}
{\it ~$^3$ Tait Institute, University of Edinburgh,\\
JCMB, KB, Mayfield Rd, Edinburgh EH9 3JZ, Scotland\\
~$^4$  Physikalisches Institut, Albert-Ludwigs-Universit\"at Freiburg,\\ 
Hermann-Herder-Stra\ss e 3, D-79104 Freiburg i. B., Germany  \\
~$^5$ Departament d'Estructura i Constituents de la Mat\`eria, 
Universitat de Barcelona,\\ Diagonal 647, E-08028 Barcelona, Spain\\
~$^6$ Institut f\"ur Theoretische Teilchenphysik und Kosmologie, RWTH Aachen University,\\ 
D-52056 Aachen, Germany\\}
\end{center}

\vspace{0.8cm}

\begin{center}
{\bf \large Abstract:}
\end{center}
We determine the strong  coupling $\alpha_s$ from a next-to-leading
order 
analysis of
processes used for the  NNPDF2.1 parton determination, which includes 
data from  
neutral and charged current deep-inelastic scattering, 
Drell-Yan and inclusive jet production. We find
$\alpha_s\lp M_Z\rp=0.1191\pm 0.0006^{\rm exp}$, where the
uncertainty includes all statistical and systematic experimental 
uncertainties, but not purely theoretical uncertainties, which are
expected to be rather larger.
We study the dependence
of the results on the dataset, by providing further determinations
based respectively
on deep-inelastic data only, and on HERA data only. The deep-inelastic 
fit gives 
the consistent result $\alpha_s\lp M_Z\rp=0.1177\pm 0.0009^{\rm exp}$, 
but the result of the HERA--only fit is only marginally consistent. We 
provide evidence that individual data subsets can have runaway directions 
due to poorly determined PDFs, thus suggesting that a global dataset 
is necessary for a reliable determination.

\clearpage

A precise knowledge of the value of  the strong coupling constant
$\alpha_s\lp M_Z\rp$~\cite{Bethke:2009jm} is necessary for accurate collider
phenomenology, such as for instance Higgs searches at
the Tevatron and the LHC~\cite{Dittmaier:2011ti}. In particular, in the 
gluon fusion channel the value of strong coupling
is one of the dominant sources of uncertainty~\cite{Demartin:2010er}.
The current PDG~\cite{Nakamura:2010zzi} value 
\be
\label{eq:bethkeav}
\alpha_s\lp M_Z\rp=0.1184 \pm 0.0007  
\ee
is taken from  Ref.~\cite{Bethke:2009jm}, where it is obtained
by combining several determinations, including some from processes
(such as the $\tau$ decay rate and the total $e^+e^-\to$~hadrons cross
section) which do not require knowledge of nucleon structure, but others (such
as deep-inelastic scattering) which do. The value of the
uncertainty
Eq.~(\ref{eq:bethkeav}) may seem overly optimistic in view of the
spread of values of available determinations and the significant
dependence on the perturbative order of some of them: as a
consequence, the use of a somewhat more conservative estimate of the
uncertainty,  such as
$\Delta\alpha_s=0.0012$ at 68\% confidence level has been
recommended~\cite{Demartin:2010er,Dittmaier:2011ti} for LHC
phenomenology.


The determination of $\alpha_s\lp M_Z\rp$ from the same wide of set of data
which is used to determine PDFs
is appealing because it simultaneously exploits
the dependence on the coupling of scaling violations as well as that
on individual hard matrix elements of the various processes under
consideration. It is thus potentially quite accurate.
On the
other hand, in such a determination the value of $\alpha_s$ is
necessarily correlated to the best-fit form of the PDFs, 
and thus subject to potential sources of
bias, such as for example an insufficiently
flexible PDF parametrization.

An example of the possible pitfalls of a simultaneous determination
of PDFs and $\alpha_s$ is highlighted by the analysis of 
Ref.~\cite{Forte:2002us}, in
which the extraction of $\alpha_s$ from BCDMS and NMC deep-inelastic
scattering data was performed using a methodology (scaling violations
of truncated moments) which avoids completely the use of parton
distributions. The result found, $\alpha_s\lp
M_Z\rp=0.124^{+0.005}_{-0.008}$, had rather different central value
and uncertainties than those obtained by direct analysis of
the same 
BCDMS ($\alpha_s\lp M_Z\rp=0.113\pm 0.005$~\cite{Virchaux:1991jc}) and
NMC ($\alpha_s\lp M_Z\rp=0.117^{+0.011}_{-0.016}$~\cite{Arneodo:1993kz}) 
data by the
respective collaborations. This
suggests that the latter results, obtained using a PDF
parametrization, were biased by it.

Here we wish to provide a determination of $\alpha_s$ exploiting the
NNPDF methodology for determining PDFs \cite{DelDebbio:2004qj,DelDebbio:2007ee,
Ball:2008by,Ball:2009mk,Ball:2010de}, which strives to avoid parametrization
bias through the use of a Monte Carlo approach combined with neural
networks as underlying unbiased interpolating functions.
Specifically, we use the latest NNPDF set,
NNPDF2.1~\cite{Ball:2011mu}, which is based on a
NLO global fit to all relevant hard scattering data, with heavy
quark mass effects included
through the so-called FONLL method~\cite{Forte:2010ta}.
NNPDF parton sets have been provided for a variety of values of
$\alpha_s$. Here we will use these sets to study the quality of the
global agreement between theory and the data used  in the PDF
determination
 as the value
of $\alpha_s$ is varied. A similar approach was used in
Ref.~\cite{Ball:2009mk} to provide a determination of the CKM matrix
element $|V_{\rm cs}|$ which turned out to be more accurate than any
determination obtained from a single experiment.

The use of NNPDF2.1 parton distributions has not only the advantage
that parametrization bias is reduced to a minimum,
but also that the same methodology can be used to analyze different
datasets, without having to retune the fitting procedure (such as, for
instance, the form of parton parametrization) according 
to the size of the dataset. This enables a
direct comparison of values of $\alpha_s$ obtained from different
subsets of data which enter the global fit, and also an analysis of
the correlation between individual datasets, individual PDFs, and the
value of $\alpha_s$. As a consequence, we will be able to address the
issue of whether deep-inelastic scattering data systematically prefer
lower values of $\alpha_s$ than hadron or $e^+e^-$
collider data.

\begin{figure}[t]
\begin{center}
\epsfig{file=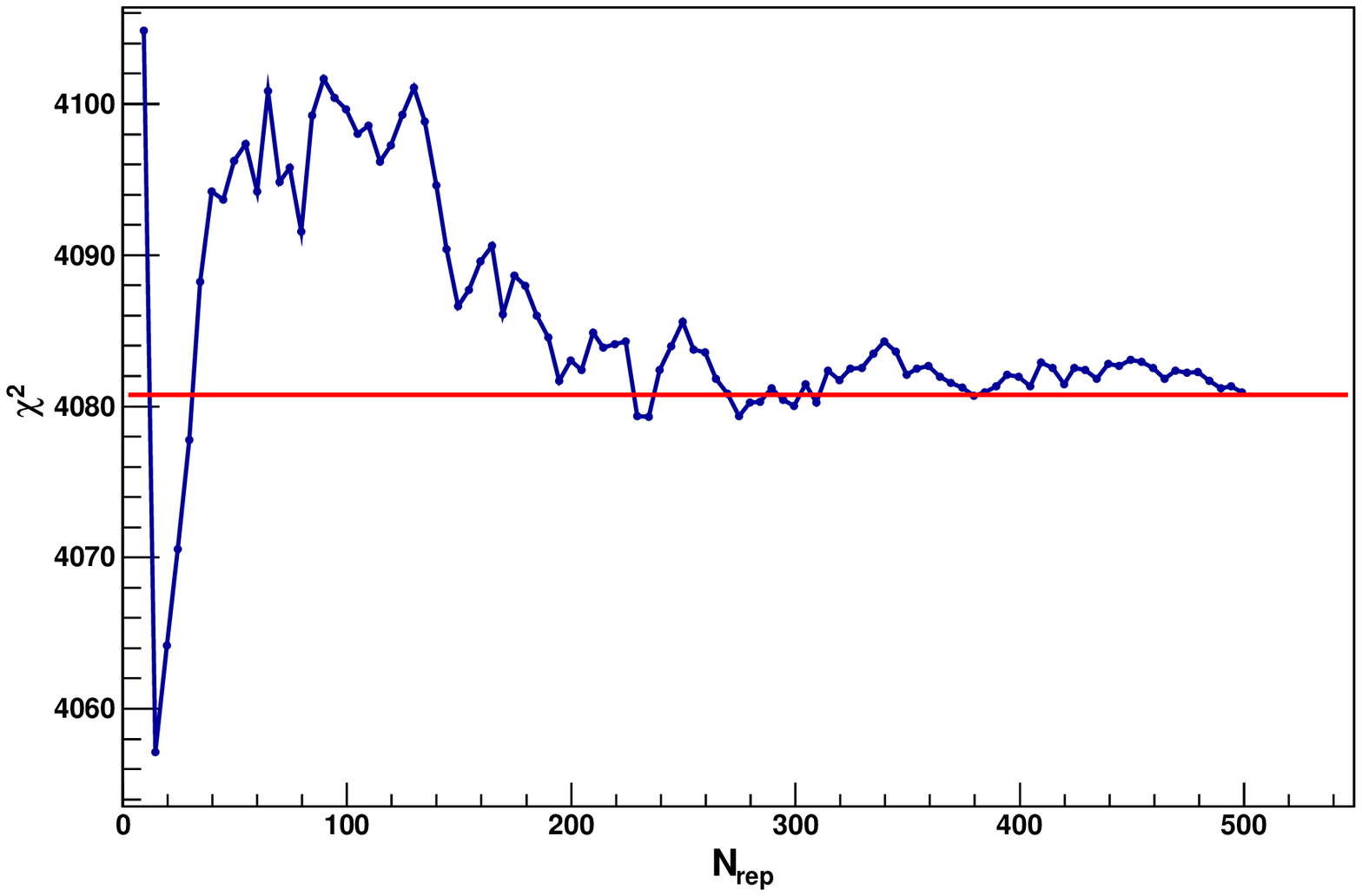,width=0.49\textwidth}
\epsfig{file=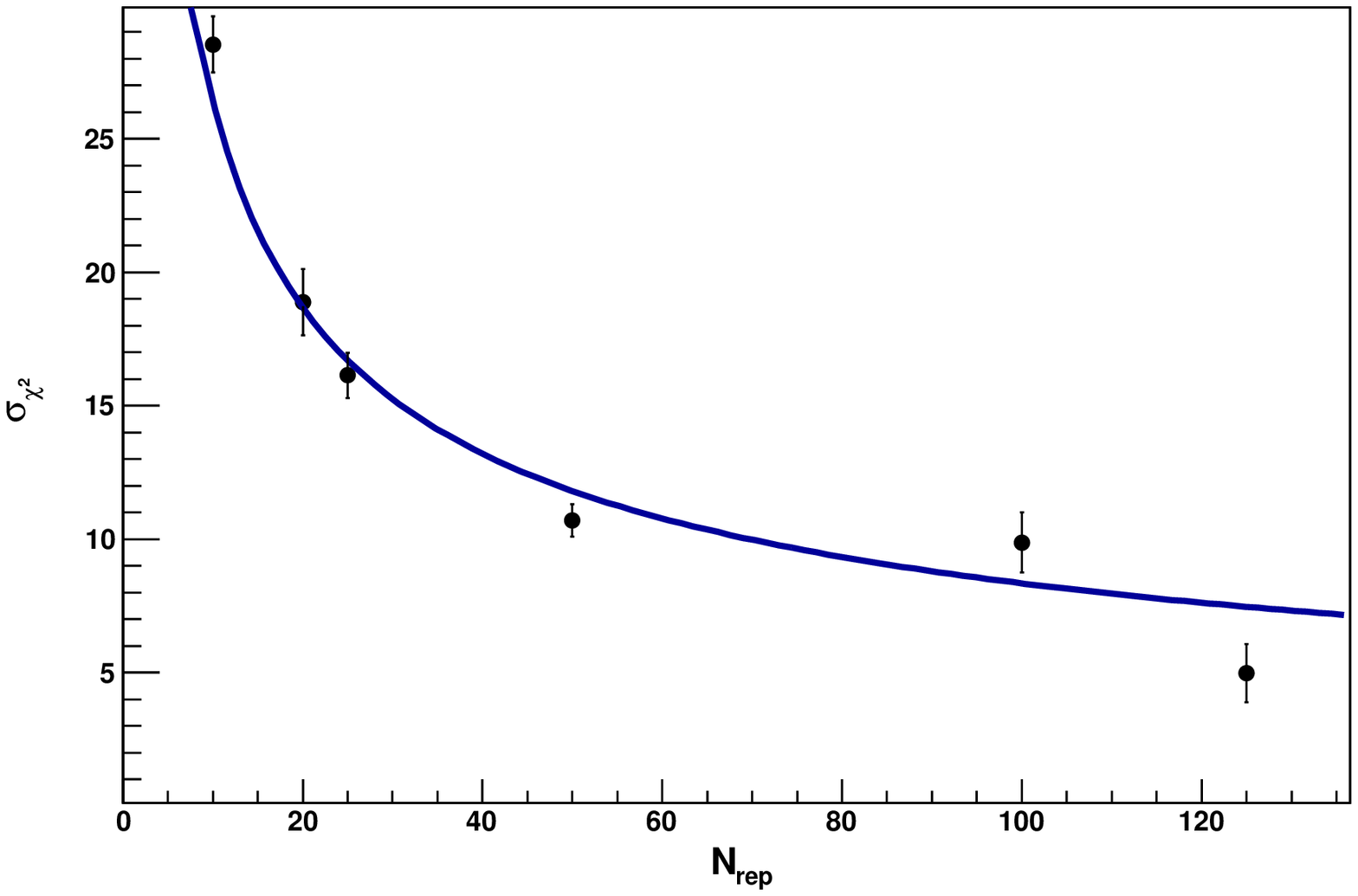,width=0.49\textwidth}
\caption{\small 
Left: The $\chi^2$ as a function 
of the number of replicas $N_{\rm rep}$ for NNPDF2.0;
the horizontal line shows the value for $N_{\rep}=500$. 
Right: The uncertainty $\sigma_{\chi^2}$ Eq.~(\ref{eq:sigchi2})
as a function of $N_{\rm rep}$ , averaged over all values of
$\alpha_s$. A fit of the form 
$A N_{\rm rep}^{-1/2}$ is also shown.
\label{fig:chi2-nrep}}
\end{center}
\end{figure}

The main difficulty  in determining $\alpha_s$ in the
NNPDF approach is that, since PDFs are delivered in the form of a
Monte Carlo sample, the quality of the fit (i.e. the $\chi^2$ of the
comparison between data and theory) is a random variable, which
only tends to a constant value in the limit in which the size of the
Monte Carlo sample tends to infinity. The typical fluctuation of
the $\chi^2$ for a single Monte Carlo replica is of the order of the
square root of the number of data points $N_{\rm dat}$, while the 
fluctuations of the average 
over a sample of  $N_{\rm rep}$ replicas decrease as 
$1/N_{\rm rep}^{1/2}$. So in order to be sensitive to variations of the 
total $\chi^2$ by a few units, as required for determination of a 
physical parameter, one needs for each value of $\alpha_s$  a number of 
replicas of the same order of magnitude as the number of 
independent data points. The total number of replicas
required is thus rather large, which makes for a rather computationally
intensive task. The value of the $\chi^2$ for a typical NNPDF
fit (with  $N_{\rm dat}=3338$) is shown as a function of $N_{\rm rep}$ in
Fig.~\ref{fig:chi2-nrep}. 

The uncertainty on the value of the $\chi^2$ due to the finite size
of the replica sample may be computed using the so-called 
bootstrap method. Namely, the
sample of $N_{\rm rep}$ replicas is divided into $N_{\rm part}$
disjoint partitions with  $\widetilde{N}_{\rm rep}= N_{\rm rep}/N_{\rm
  part}$ replicas each. The variance of the $\chi^2$ for the full
$N_{\rm rep}$ replica sample is then found from the variance of the
$N_{\rm part}$ values
$\widetilde{\chi}^2$  of each replica  subsample according to
\be
\lp \sigma_{\chi^2}\rp^2
\equiv \frac{1}{N_{\rm part}}\lc \frac{1}{N_{\rm part}}
\sum_{k=1}^{N_{\rm part}} \lp \widetilde{\chi}^2_{k}\rp^2 -
\lp \frac{1}{N_{\rm part}}
\sum_{k=1}^{N_{\rm part}} \widetilde{\chi}^2_{k} \rp^2  \rc \ .
\label{eq:sigchi2}
\ee
The value of $\sigma_{\chi^2}$, averaged over all the (eleven) values of
$\alpha_s$ to be considered, is displayed in
Fig.~\ref{fig:chi2-nrep}. A fit of the form
$AN_{\rm rep}^{-1/2}$, also shown in
Fig.~\ref{fig:chi2-nrep}, shows that the expected decrease of the
fluctuations with  $1/N_{\rm rep}^{1/2}$ is borne out by
the data.

The determination of $\alpha_s$ is performed by simply
using  a wide enough  PDF replica set for $N_{\alpha_s}$ fixed values of 
$\alpha_s$ to compute for each value of $\alpha_s$
the values of  $\chi^2$ and its uncertainty $\sigma_{\chi^2}$
Eq.~(\ref{eq:sigchi2}), and then fitting a parabola to the $\chi^2$
viewed as a function of $\alpha_s$.
The quality 
of the parabolic fit is then determined by evaluating the corresponding 
$\chi^2_{\rm par}/N_{\rm dof}$, with $N_{\rm dof}=N_{\alpha_s}-3$: a 
reasonable value of $\chi^2_{\rm par}/N_{\rm dof}$ may be used to 
confirm that the parabolic approximation to
$\chi^2(\alpha_s)$ is adequate in the range of $\alpha_s$ under
investigation. The minimum of the parabola then provides
the best-fit value of $\alpha_s$ while the $\Delta\chi^2=1$
range gives the uncertainty on it at a 68\% confidence level.
The further uncertainty due to the finite size
of the replica sample is determined by error propagation of  $\sigma_{\chi^2}$
Eq.~(\ref{eq:sigchi2}) on the position of the minimum of the parabola.   

\begin{figure}[t]
\begin{center}
\epsfig{file=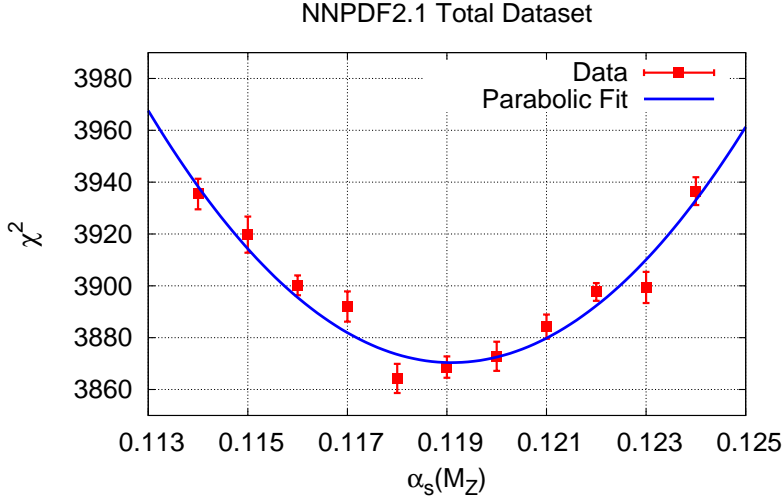,width=0.70\textwidth}
\caption{\small 
The $\chi^2$ as a function of $\alpha_s(M_Z)$ for the 
NNPDF2.1 global fit. The statistical
uncertainties in the $\chi^2$ for each
value of $\alpha_s$ have been determined
from Eq.~(\ref{eq:sigchi2}).
The solid line is the result
of a parabolic fit. 
\label{fig:21total} }
\end{center}
\end{figure}

\begin{figure}[t]
\begin{center}
\epsfig{file=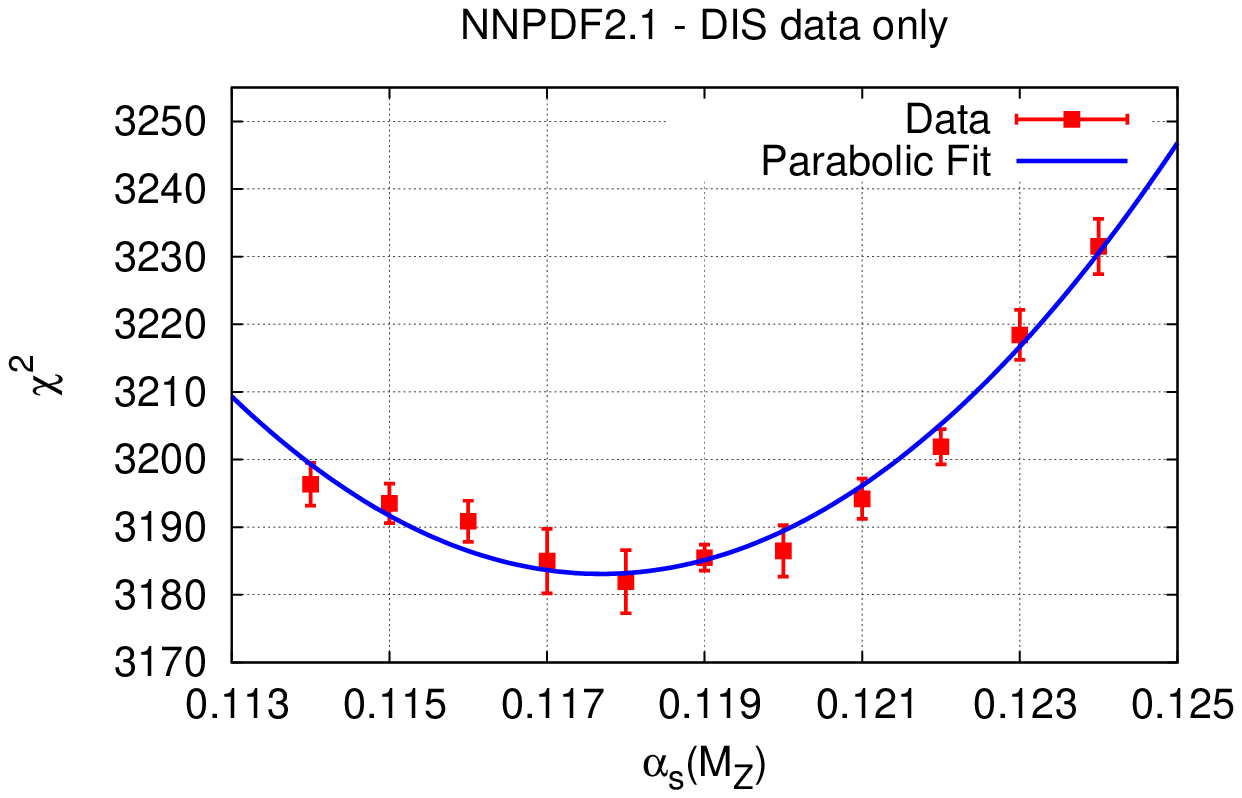,width=0.48\textwidth}
\epsfig{file=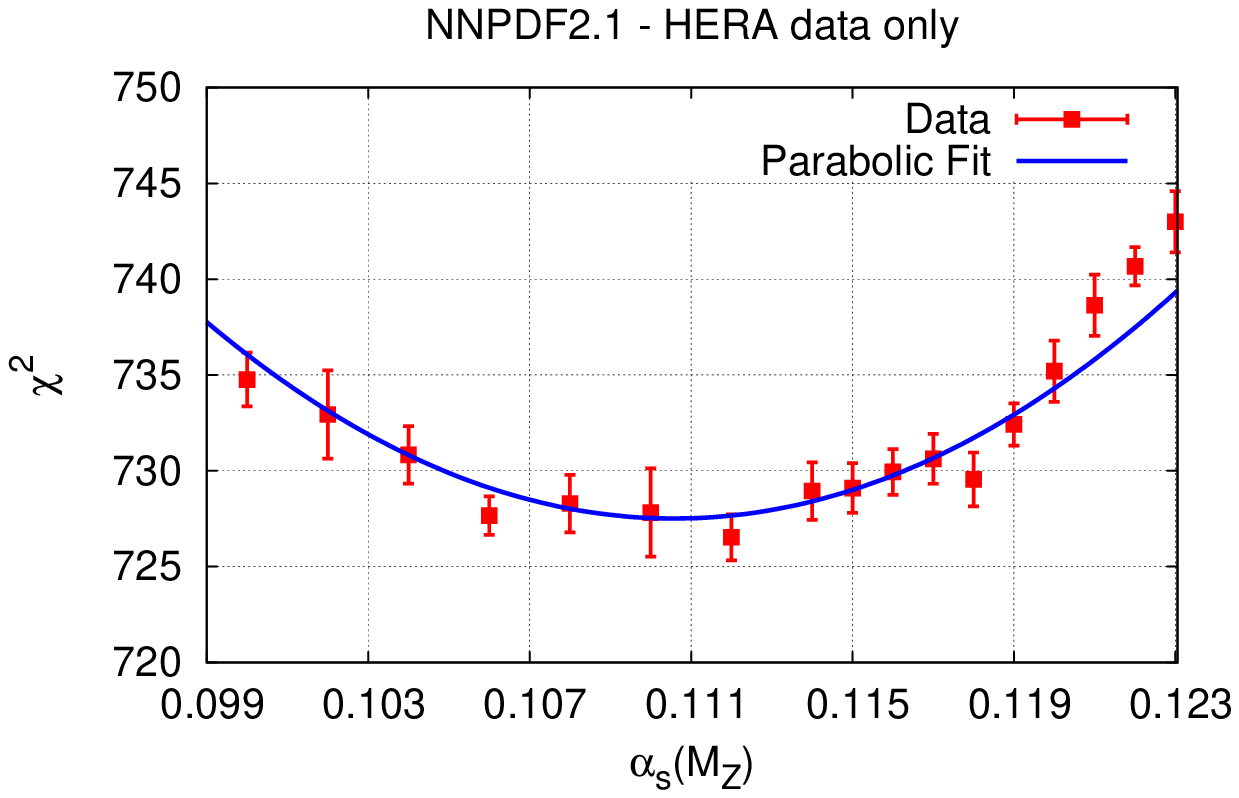,width=0.48\textwidth}
\caption{\small 
Same as Fig.~\ref{fig:21total} but for a fit to DIS data only (left) 
and to HERA data only (right). 
\label{fig:21red} }
\end{center}
\end{figure}

We now turn to results. 
For each 
value of $\alpha_s$ we use a sample of at least $N_{\rm rep}=500$
replicas, with bigger samples of $N_{\rm rep}=1000$ replicas used for
more sparse equally--spaced values in order to increase accuracy. 
For the fit to HERA data only, the range of values
considered has been enlarged in order to ensure that the location of
the minimum is approximately at the center of the region of $\alpha_s$
which is being explored, and also because in this case the sensitivity
to $\alpha_s$ is weaker due to the much smaller size of the data sample.
The values of $\alpha_s$ and numbers of replicas used in each case are
summarized in Table~\ref{tab:asnrep}.

\begin{table}
\tiny
\begin{center}
\begin{tabular}{|c|c|}
\hline
\multicolumn{2}{|c|}{2.1 global} \\
$\alpha_s\lp M_Z\rp$ & $N_{\rm rep}$ \\
\hline
\hline
0.114  &  500 \\
0.115  &  500 \\
0.116  &  1000 \\
0.117  &  500 \\
0.118  &  500 \\
0.119  &  1000 \\
0.120  &  500 \\
0.121  &  500 \\
0.122  &  1000 \\
0.123  &  500 \\
0.124  &  500 \\
\hline
\end{tabular}
\begin{tabular}{|c|c|}
\hline
 \multicolumn{2}{|c|}{2.1 DIS--only} \\
$\alpha_s\lp M_Z\rp$ & $N_{\rm rep}$ \\
\hline
\hline
0.114  &  500 \\
0.115  &  500 \\
0.116  &  1000 \\
0.117  &  500 \\
0.118  &  500 \\
0.119  &  1000 \\
0.120  &  500 \\
0.121  &  500 \\
0.122  &  1000 \\
0.123  &  500 \\
0.124  &  500 \\
\hline
\end{tabular}
\begin{tabular}{|c|c|}
\hline
\multicolumn{2}{|c|}{2.1 HERA--only} \\
$\alpha_s\lp M_Z\rp$ & $N_{\rm rep}$ \\
\hline
\hline
0.100  &  1000 \\
0.102  &  500 \\
0.104  &  500 \\
0.106  &  1000 \\
0.108  &  500 \\
0.110  &  500 \\
0.112  &  1000 \\
0.114  &  500 \\
0.115  &  500 \\
0.116  &  1000 \\
0.117  &  500 \\
0.118  &  500 \\
0.119  &  1000 \\
0.120  &  500 \\
0.121  &  500 \\
0.122  &  1000 \\
0.123  &  500 \\
0.124  &  500 \\
\hline
\end{tabular}
\end{center}
\caption{\small The values of $\alpha_s\lp M_Z\rp$ and the
number of replicas $N_{\rm rep}$ used in each case  for various
determinations of $\alpha_s\lp M_Z\rp$. \label{tab:asnrep}}
\end{table}

The parabolic profile of
$\chi^2$ as a function of $\alpha_s(M_Z)$ is shown
in Fig.~\ref{fig:21total} for the  NNPDF2.1 global fit.
Analogous results for fits to DIS data only and to HERA data only are
shown in  Fig.~\ref{fig:21red}. 
The corresponding 
values and uncertainties of $\alpha_s\lp M_Z\rp$ are collected
Table~\ref{tab:alphasfit}.
In each case, we denote with ``exp''
the uncertainty from the $\Delta\chi^2=1$ range and with ``proc'' the
propagated ``procedural'' uncertainty, due the finite size of the replica sample.
The quality of the parabolic fit is
also shown in each case.

\begin{table}[t]
\centering
\begin{tabular}{|c|c|c|}
\hline
  & $\alpha_s\lp M_Z\rp$  & $\chi^2_{\rm par}/N_{\rm dof}$  \\
\hline
\hline
{\bf NNPDF2.1} &  ${\bf 0.1191 \pm 0.0006^{\rm exp} \pm 0.0001^{\rm
    proc} } $ & {\bf {1.6}} \\
\hline
NNPDF2.1 DIS--only & $0.1178 \pm 0.0009^{\rm exp} \pm 0.0002^{\rm proc}  $ & 0.7 \\
NNPDF2.1 HERA--only & $0.1101 \pm 0.0033^{\rm exp} \pm 0.0003^{\rm proc}  $ & 0.7 \\
\hline
NNPDF2.1 {\it red.} & $0.1191 \pm 0.0006^{\rm exp} \pm 0.0001^{\rm proc} $ & 1.5 \\
NNPDF2.1 DIS--only  {\it red.}& $0.1177 \pm 0.0009^{\rm exp} \pm 0.0002^{\rm proc} $ & 0.5 \\
NNPDF2.1 HERA--only  {\it red.} & $0.1103 \pm 0.0032^{\rm exp} \pm 0.0004^{\rm proc} $ & 1.1 \\
\hline
NNPDF2.0 &  $0.1168 \pm 0.0007^{\rm exp}  \pm 0.0001^{\rm proc}  $ 
& 0.4 \\
NNPDF2.0 DIS--only &  $0.1145 \pm 0.0010^{\rm exp} \pm 0.0003^{\rm proc} $
& 1.4. \\
\hline

\end{tabular}
\caption{\small Values of $\alpha_s\lp M_Z\rp$ and associated
  uncertainties. All uncertainties shown are 68\% confidence levels,
  with the experimental uncertainty obtained by requiring 
$\Delta\chi^2=1$
  about the minimum, and the procedural uncertainty from propagation
  of $\sigma_{\chi^2}$
Eq.~(\ref{eq:sigchi2}) due to finite size of the replica sample.
The quality of the parabolic fit as measured by  
$\chi^2_{\rm par}/N_{\rm dof}$ is also shown in each case. For the
global, DIS--only and HERA--only fits (first three rows), the maximum
number of replicas, given in Tab.~\ref{tab:asnrep}, has been used. The
three reduced replica  fits (subsequent three rows) only differ from
these because of the use of $N_{\rm rep}=500$ for all $\alpha_s$
values. The 
 NNPDF2.0 fits of the last two rows also have $N_{\rm rep}=500$ always. 
\label{tab:alphasfit} }
\end{table}

\begin{figure}[t]
\begin{center}
\epsfig{file=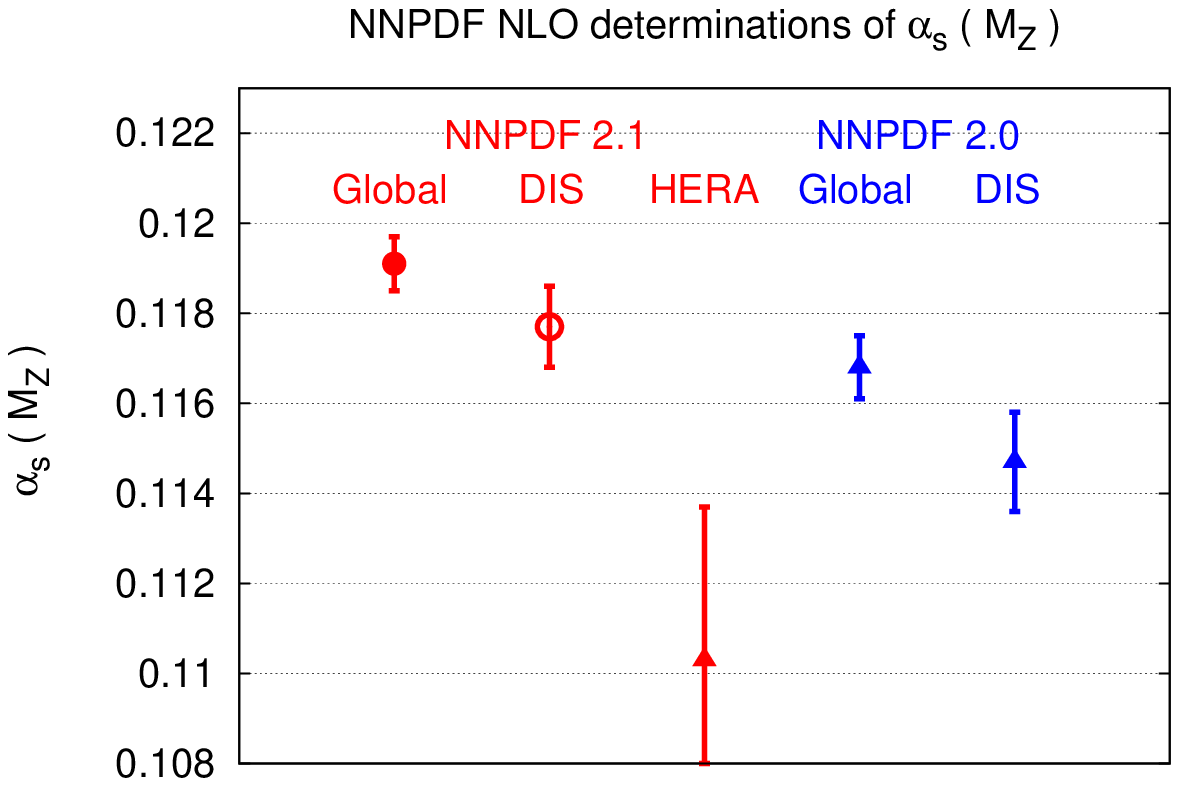,width=0.49\textwidth}
\epsfig{file=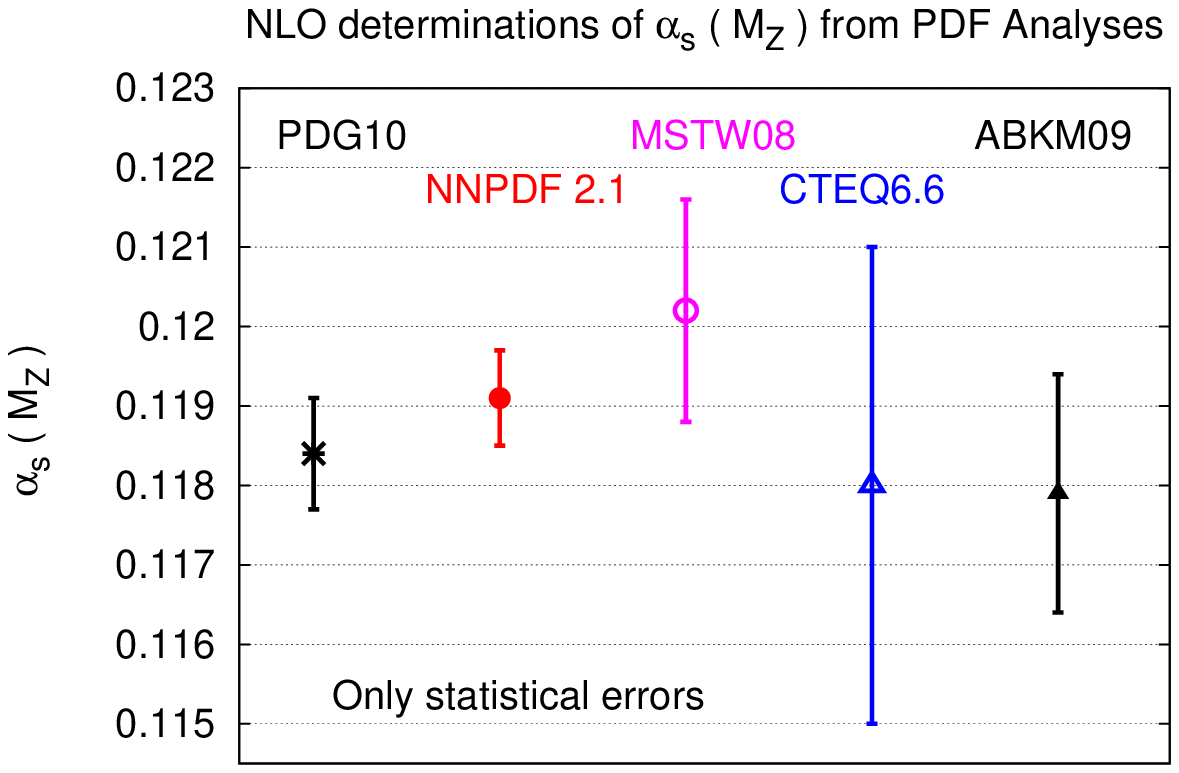,width=0.49\textwidth}
\end{center}
\caption{\small Left: Graphical representation of the values  of
$\alpha_s\lp M_Z\rp$ of Table~\ref{tab:alphasfit} (reduced replica
  fits not included).  Uncertainties have 
  been added in quadrature). Right:
comparison to other recent determinations of $\alpha_s\lp M_Z\rp$
from NLO PDF analysis. The PDG value of Ref.~\cite{Bethke:2009jm}
is also shown.
\label{fig:alphas-res-comp}}
\end{figure}

The procedural uncertainties in Tab.~\ref{tab:alphasfit} are all very
small. This implies that the
best-fit value of $\alpha_s$ is already approximately independent 
of   the size of the replica
sample, and a further increase in the number of replicas is not
necessary. 
However, it is useful to check this independence 
explicitly by varying the number
of replicas, in order to make sure that the finite-size uncertainty
has been determined correctly. To this purpose,
 we have repeated the three NNPDF2.1 determination of $\alpha_s$
with   $N_{\rm rep}=500$ for all values of
$\alpha_s$. Results are also collected in Table~\ref{tab:alphasfit} and indeed show
excellent stability: 
the change in value of $\alpha_s$ is always
smaller than the procedural uncertainty as the number of replicas is decreased. 
The $\chi^2$ values of the parabolic fit should follow a
$\chi^2$ distribution with  $N_{\rm dof}=8$ degrees of freedom 
for the global and DIS fits, and
$N_{\rm dof}=12$ for the HERA only fit. The standard deviation of
$\chi^2_{\rm par}/N_{\rm dof}$ is thus expected to be
of order $0.5$, as indeed observed.
 We have finally checked that excluding the points at the 
edge of the fit, and adding extra parameters to the fit, has no 
significant effect on the results, and in particular it does not
improve the quality of the parabolic fit.

Our results for $\alpha_s$ are displayed graphically in 
Fig.~\ref{fig:alphas-res-comp}. 
Our best-fit value of $\alpha_s$ is in good agreement with the 
PDG value Eq.~(\ref{eq:bethkeav}), and has a surprisingly small
experimental uncertainty. The experimental uncertainty increases as the
size of the dataset is reduced, as it ought to. 
We see no evidence that DIS data prefer a 
significantly lower value of $\alpha_s$: the difference between values
of the global and DIS-only determinations is of order of one $\sigma$,
and thus entirely compatible with statistical fluctuations.
Interestingly, the value found using HERA
data only is much smaller, even though, because of the considerable
(almost sixfold) increase in statistical uncertainty it is still
less than three $\sigma$ from the global fit. The fact
that HERA data prefer a lower value of $\alpha_s$ may be related to
the deviations between HERA data and the predicted NLO scaling
violations which was observed in
Refs.~\cite{Caola:2009iy,Caola:2010cy} for the smallest $x$ and $Q^2$
HERA data: these  may be affected by
small $x$ resummation or saturation effects. As shown there, 
scaling violations in this region are weaker than
predicted from the behaviour observed in other kinematic regions, and
thus would tend to bias the value of $\alpha_s$ downwards. A dedicated
 analysis would be required to prove conclusively that this is the case.

We make no attempt here to estimate theoretical uncertainties in our
fit. Uncertainties due to inefficiencies of the global PDF fit (such
as, for example, any residual bias related to parton distributions) should
show up in the behaviour of the $\chi^2$ as a function of $\alpha_s$,
either as point-to-point fluctuations or as a systematic deviation
from the underlying unbiased quadratic behaviour (if they are
correlated to the value of $\alpha_s$). The good quality of the
parabolic fit suggests that these uncertainties are small, and thus
that our uncertainty is an accurate assessment of the total
uncertainty due to the statistical and systematic uncertainties in the 
experimental data. On top of these, however, there will be genuine
theoretical uncertainties related to the theory used in the
computation of the various processes under investigation. Of these,
the main ones are likely to be related to NNLO and higher QCD
corrections (and possibly resummation of higher order QCD corrections
in some kinematic regions), and to the treatment of heavy quark mass
effects. These uncertainties are presumably of similar size here as in
other determinations of $\alpha_s$ based on the same QCD processes;
whereas they were studied systematically in older $\alpha_s$
determinations (such as Ref.~\cite{Ball:1995qd}),
they have not been assessed for any of the more recent determinations.
In  cases in which both NLO and NNLO determinations are
available, such as Ref.~\cite{Martin:2009bu, Alekhin:2009ni}, a
sizable downward shift of the best-fit value, of order of several
percentage points, has been observed when going from NLO to NNLO. It
will be interesting to see whether such an effect is also present in
the NNPDF approach.

\begin{figure}[t]
\begin{center}
\epsfig{file=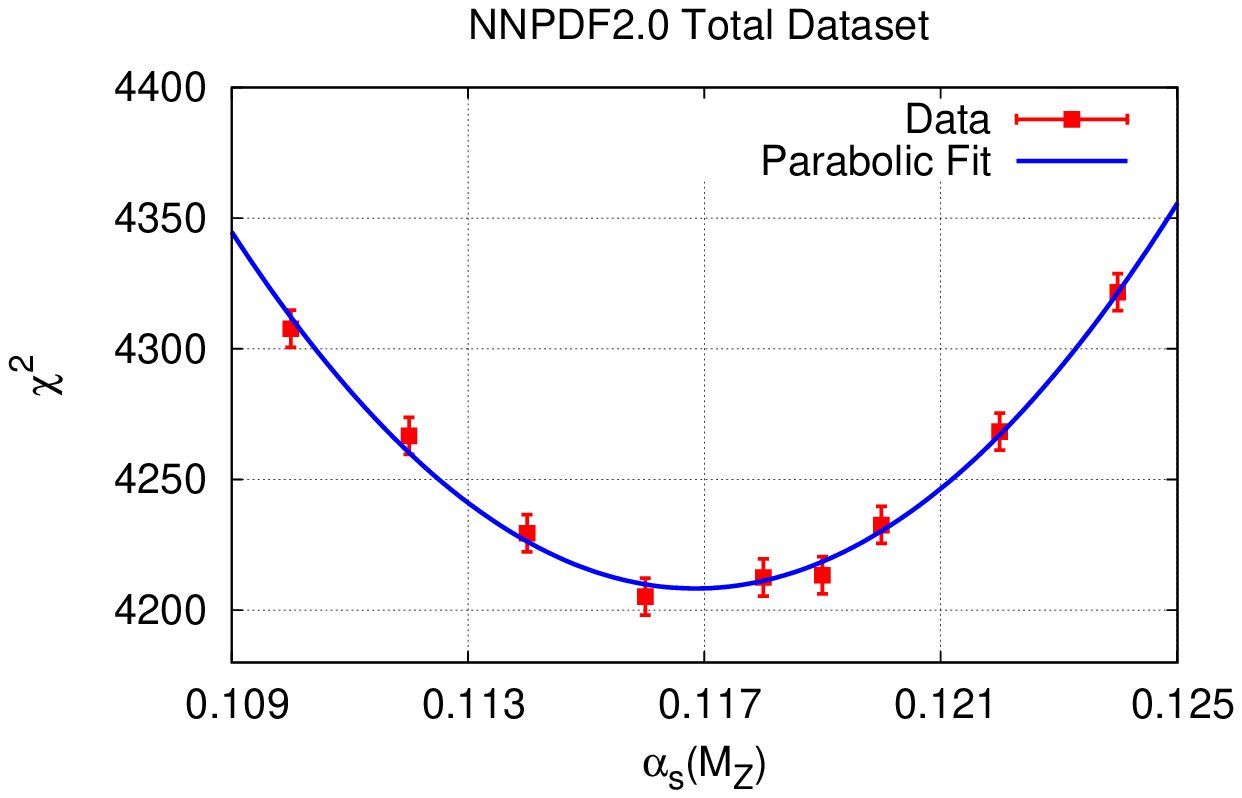,width=0.48\textwidth}
\epsfig{file=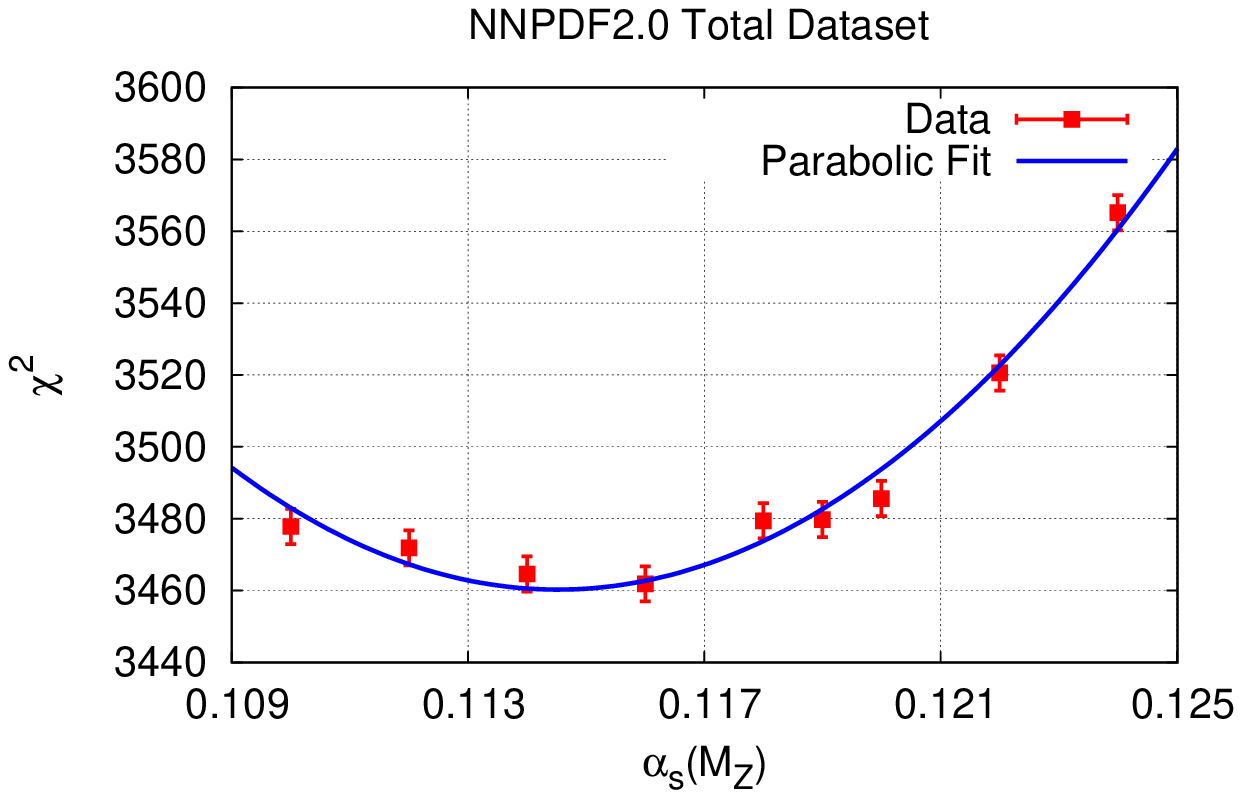,width=0.48\textwidth}
\caption{\small 
Same as Fig.~\ref{fig:21total} but for the NNPDF2.0  global fit
(left) and NNPDF2.0 DIS only fit (right). 
\label{fig:nnpdf20fit} }
\end{center}
\end{figure}
As a very crude estimate of the order of magnitude of effects related
to heavy quark masses, we have repeated the fit to the global
dataset and that to DIS data using the NNPDF2.0~\cite{Ball:2010de} PDF
set, which is based on a zero-mass variable flavour number scheme, in
which all heavy quark masses are neglected. The results are also given
in Table~\ref{tab:alphasfit} and shown in 
Fig.~\ref{fig:alphas-res-comp}, while the corresponding parabolic fits
are displayed in Fig~\ref{fig:nnpdf20fit}. For these fits, $N_{\rm rep}=500$
replicas are used for all values of $\alpha_s$.
Neglecting heavy quark mass effects induces a
significant downward shift in $\alpha_s$. If one were to conservatively
estimate the uncertainty due to heavy quark mass effects
as the difference between
the NNPDF2.1 and NNPDF2.0 results one would get, for the global fit,
$\Delta\alpha_s^{\rm hq}\approx 0.002$. In the fit to DIS data only
the shift is larger since the hadronic data
are unaffected by the treatment of heavy quark mass effects.
It is likely that the order of magnitude of uncertainties related 
to higher order corrections is comparable, so that the 
theoretical uncertainty is very likely to be the dominant one.

Our result using the standard (global) NNPDF2.1 set 
is compared in Fig.~\ref{fig:alphas-res-comp} to other recent NLO
determinations of $\alpha_s$ which rely respectively on the
MSTW~\cite{Martin:2009bu}, CTEQ~\cite{Lai:2010nw} and
ABKM~\cite{Alekhin:2009ni} PDF sets (ABKM and MSTW also provide
determinations at NNLO). The MSTW and ABKM groups
perform a simultaneous fit of PDFs and the strong coupling, thus
obtaining a correlated Hessian matrix which mixes the PDF parameters
with $\alpha_s$, while CTEQ simply studies the dependence of the fit
quality on $\alpha_s$ as is done here (see Sect.~3 of
Ref.~\cite{Alekhin:2011sk}). The equivalence of the
$\alpha_s$ uncertainty obtained from either method
is explicitly shown in Ref.~\cite{Lai:2010nw}. The dataset
on which the CTEQ and MSTW determinations are based is very similar to
our own, while ABKM use a smaller dataset, which in particular does
not include collider jet and vector boson production data. 
All these determinations are in agreement with each other within
uncertainties. The rather larger statistical uncertainties found by
CTEQ and MSTW can be understood as a consequence of the fact that
these groups~\cite{Martin:2009bu,Lai:2010nw} use a
tolerance~\cite{Pumplin:2001ct}  criterion to obtain 68\% confidence
levels, based on a substantial rescaling of the uncertainty ranges
in parameter space. This is not necessary in our approach because,
once the $\chi^2$ is treated as a random variable, its fluctuations can
be studied (and in particular kept under control) by a suitable choice
of the size of the Monte Carlo sample, as discussed above.

\begin{figure}[t]
\begin{center}
\epsfig{file=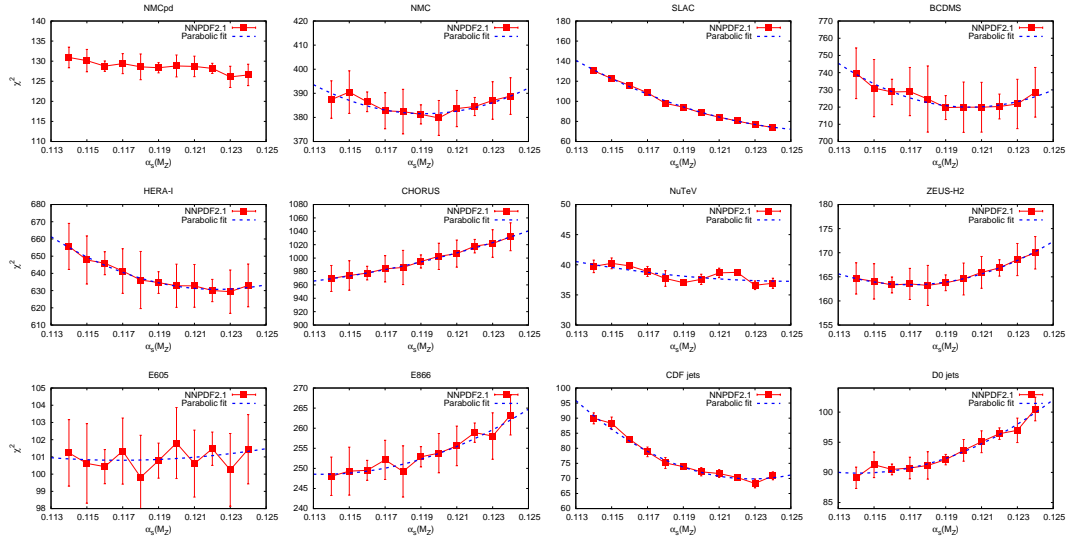,width=\textwidth}
\caption{\small 
The $\chi^2$ profiles for the individual experiments in the
NNPDF2.1 global fit together with the results of the
corresponding parabolic fits to $\alpha_s$. 
The uncertainties due to the finite size of the replica sample are
shown on each value.
\label{fig:alphasfit-exps}}
\end{center}
\end{figure}

\begin{figure}[h]
\begin{center}
\epsfig{file=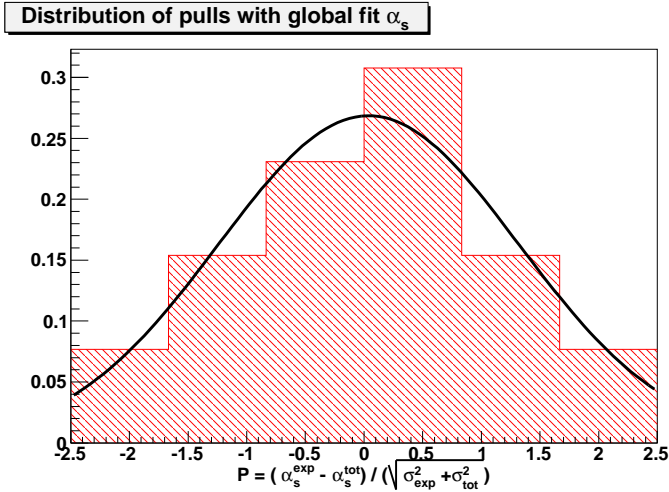,width=0.65\textwidth}
\caption{\small 
Distribution of pulls, Eq.~(\ref{eq:pull}),
for the value of $\alpha_s$
preferred by the individual experiments included
in the global fit. These pulls have
been summarized in Table~\ref{fig:alphasfit-exps}).
\label{fig:alphas-histo}}
\end{center}
\end{figure}

\begin{table}
\begin{center}
\small
\begin{tabular}{|c|c|c|}
\hline
Experiment &     $\alpha_s^i\pm\sigma_{\alpha_s}^i$  &         $P_i$\\
\hline
\hline
NMCp  & $0.1192 \pm 0.0018$  &   -0.05 \\
BCDMS & $0.1204 \pm 0.0015$  &  -0.78 \\
HERA-I & $0.1223 \pm 0.0018$ &  -1.65 \\
ZEUS-H2 & $0.1170 \pm 0.0027$ &   0.75 \\
NuTeV & $0.1252 \pm 0.0068$   & -0.89 \\
ZEUSF2C & $0.1144 \pm 0.0060$ &  0.77 \\
\hline 
E605 & $0.1168 \pm 0.0100$    &  0.22  \\
E866 & $0.1135 \pm 0.0029$    &  1.87  \\
CDFWASY & $0.1181 \pm 0.006$  &  0.16  \\
CDFZRAP & $0.1150 \pm 0.0034$ &  1.18  \\
D0ZRAP & $0.1227 \pm 0.0067$  &  -0.53  \\
\hline
CDFR2KT & $0.1228 \pm 0.0021$ & -1.67   \\
D0R2CON & $0.1141 \pm 0.0031$ &  1.57   \\
\hline
\end{tabular}
\end{center}
\caption{\small The pulls $P_i$ Eq.~(\ref{eq:pull}) for 
individual experiment included in the 
NNPDF2.1 global fit case, computed  for each experiment
which has a minimum in the range considered.
 \label{tab:as-pulls} }
\end{table}

This conclusion may be cross-checked using a variant of the method
suggested in Ref.~\cite{Pumplin:2009sc}, namely, by checking whether the
distribution of results obtained from individual datasets follows a
gaussian distribution and determining the width of this distribution.
To this purpose, we have performed a parabolic fit to the
 $\chi^2$ profile for each experiment entering in the global
NNPDF2.1 determination. Results are displayed in 
Fig.~\ref{fig:alphasfit-exps} together with
the uncertainties due to the finite size of the replica sample,
determined as above. These uncertainties are clearly much larger than
the point-to-point fluctuations of the individual $\chi^2$ values for
each experiment, due to the fact that the latter, being determined
from a global fit, are strongly correlated with each other. 
The  
value of $\alpha_s$ and its statistical
uncertainty for each experiment are determined by performing a
parabolic fit to each $\chi^2$ profile: results are collected   
in Table~\ref{tab:as-pulls}, for all experiments for which there is a
minimum in the fitted range. This is not the case for the 
NMCratio, SLAC, CHORUS, H1F2C and FLH108 data.

The distribution of results can be studied defining the  
pull 
\be
\label{eq:pull}
P_i\equiv \frac{\alpha_s^i\lp M_Z\rp-\alpha_s^{\rm tot}\lp M_Z\rp}{\sqrt{\sigma^{i,2}_{\alpha_s}+
\sigma^{\rm tot,2}_{\alpha_s}}},
\ee
where  $\alpha_s^i\lp M_Z\rp$  is the best fit value for the $i$--th experiment
and $\sigma^i_{\alpha_s}$ the associated statistical uncertainty,
obtained from
the $\Delta\chi^2=1$ rule. The pulls are summarized
in Table~\ref{tab:as-pulls}  
and displayed graphically in Fig.~\ref{fig:alphas-histo}. A 
gaussian 
fit to the distribution of pulls is  performed, and also displayed in
Fig.~\ref{fig:alphas-histo}. 
The gaussian fit is in good agreement with the histogram data
with mean $\langle P\rangle =0.04$ and standard deviation $\sigma_P = 1.3$. The
standard deviation would
be further reduced somewhat if finite-size uncertainties were included;
this however would require a lengthy correlation analysis. 
We conclude that the value of the tolerance required to get a
perfectly gaussian distribution of pulls is smaller than $1.3$ --- a
value which is clearly compatible with a statistical fluctuation.

\begin{figure}
\begin{center}
\epsfig{file=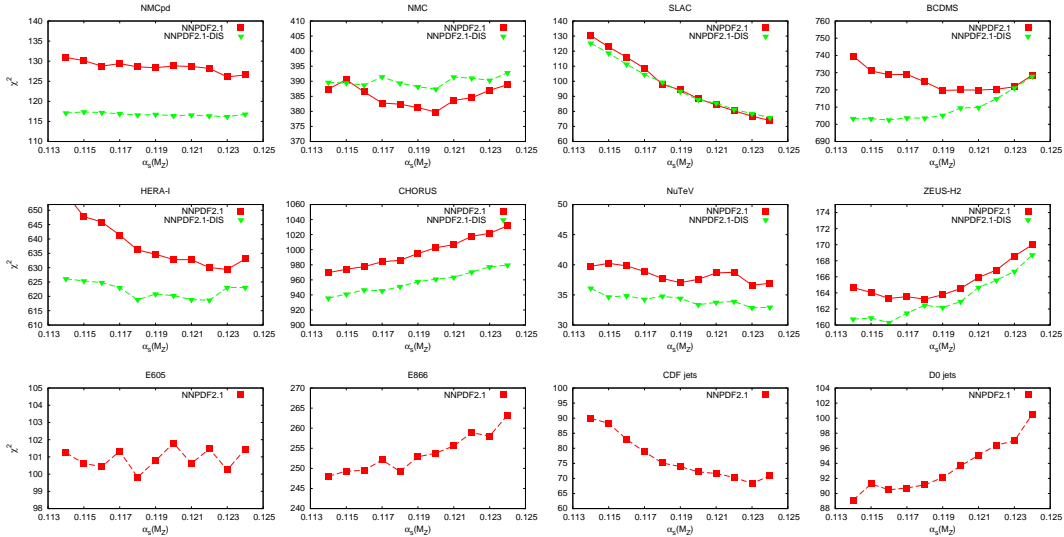,width=\textwidth}
\caption{\small 
Comparison of the $\chi^2$ profiles for the global NNPDF2.1 fit (same
as in Fig.~\ref{fig:alphasfit-exps}; red, solid curves) to those
determined for 
the DIS-only NNPDF2.1 fit
(green, dashed curves).
\label{fig:alphasfit-exps-nnpdf21} }
\end{center}
\end{figure}

Finally, we exploit the fact that in our approach a single 
procedure can be used to obtain PDFs from datasets of different size,
in order
to study the issue of whether (and why) different values of $\alpha_s$
may be preferred by different datasets.
In Fig.~\ref{fig:alphasfit-exps-nnpdf21} we compare  the $\chi^2$
profiles for the global NNPDF2.1 fit, already shown in
Fig.~\ref{fig:alphasfit-exps}, to the same quantities determined for
the fit to DIS data only. 
The behaviour of the fit quality for BCDMS data is particularly
interesting: these data have been repeatedly found~\cite{Virchaux:1991jc,Martin:2009bu,Martin2011} to
prefer a relatively low value of $\alpha_s$ (in particular, lower than
Eq.~(\ref{eq:bethkeav})). It turns out that the $\chi^2$ profile for
these data is rather different according to whether one is looking at
PDFs determined using DIS data only (green, lower curve in 
Fig.~\ref{fig:alphasfit-exps-nnpdf21}) or a  global fit
(red, higher curve in in
Fig.~\ref{fig:alphasfit-exps-nnpdf21}). Indeed, in the
DIS-only case these data indeed seem to prefer a lower value of
$\alpha_s$, but this is no longer the case in the global fit. This
suggests that as $\alpha_s$ is lowered, the quality of the fit to
BCDMS data can be improved by changing the PDFs in a way which is
allowed by DIS data, but which is not compatible with other data in
the global fit. In other words, there is a runaway direction in the
space of PDF parameters along which the $\chi^2$ for BCDMS data
decreases as $\alpha_s$ is lowered, which is forbidden 
in the global fit because then the $\chi^2$ for some other
dataset would increase.

\begin{figure}
\begin{center}
\epsfig{file=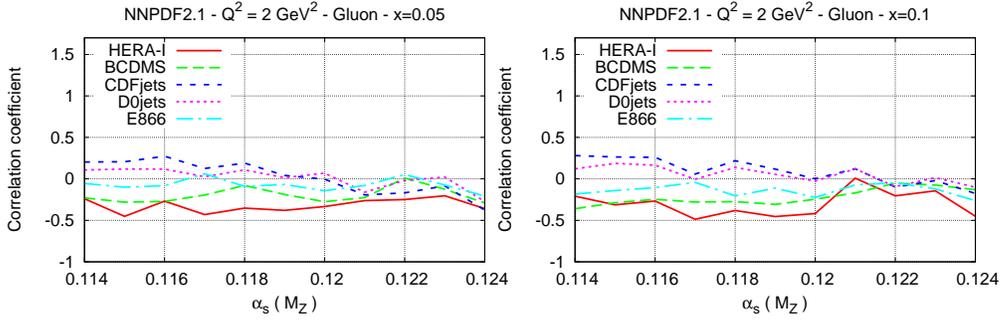,width=\textwidth}
\caption{\small 
Correlation between the $\chi^2$ and the input gluon as 
a function of $\alpha_s\lp M_Z\rp$ for
$x=0.05$ (left) and $x=0.1$ (right) for the NNPDF2.1 global PDF set.
\label{fig:corr-pdfs-chi2} }
\end{center}
\end{figure}

The situation can be further elucidated by studying the
correlation~\cite{Ball:2011mu} 
between parton distributions and the value of the $\chi^2$ for individual 
experiments. 
The presence of a nonvanishing correlation means that, at
the best fit, the $\chi^2$ for that experiment is not stationary, i.e.
it can be lowered or raised
 by changing the given PDF. Correlations of
opposite sign for different experiments then mean that these experiments
are pulling the PDF in opposite directions. The correlation
coefficients are shown for the gluon PDF
as a function of $\alpha_s$ 
in Fig.~\ref{fig:corr-pdfs-chi2} for a pair of values of $x$, for the
global NNPDF2.1 PDF set (all computed from a set of 
$N_{\rm rep}=500$ replicas).
It is apparent that  while for larger values of $\alpha_s\sim 0.120$
correlations are small and with the same sign, as $\alpha_s$ is
lowered correlations become larger, with opposite sign for jet and DIS
(HERA and BCDMS) 
experiments (with the Drell-Yan experiment E866 showing no significant
correlation). This means that indeed, as suggested  above, for low
$\alpha_s$  DIS and jet
experiments pull the gluon in opposite directions, while they become
more consistent for larger $\alpha_s$: hence, a determination of
$\alpha_s$  including DIS data
only can easily be biased. The fact that BCDMS data prefer a lower
value of $\alpha_s$ in a DIS-only fit, but not if the gluon is
constrained by jet data was also found recently in
Ref.~\cite{Martin2011}, in the context of the MSTW08 parton
determination. However, in that case the BCDMS data were also found to
significantly bias downwards the value of $\alpha_s$ of the DIS fit,
perhaps due to the fact that the MSTW gluon parametrization, though
more flexible than that of other groups, is still less flexible than
that of present analysis.

The fact that runaway directions for the $\chi^2$ 
may appear in the joint $\alpha_s$-gluon space 
can be understood by noting that in DIS the gluon is
determined by scaling violations, hence a smaller value of $\alpha_s$
can be partially compensated by a larger gluon and conversely.  
However, the jet
cross section pins down the size of the gluon (at the rather larger
scale of the jet data) thereby quenching this potential instability.
Hence, we conclude that even though in our fit 
the DIS-only value of $\alpha_s$ 
is not significantly smaller than that for
the global fit (possibly due to the great flexibility of the
functional form of our PDFs), a fit to DIS data, and specifically to
BCDMS data, has a potential
instability in the direction of lower values of $\alpha_s$ which is only
kept under control by the inclusion of jet data.

We conclude that a reliable $\alpha_s$ determination, with
surprisingly small statistical uncertainty, can be obtained by
a combined analysis of a wide set of data which simultaneously depend on the
value of the strong coupling and the parton distributions.
Theoretical uncertainties are likely to be dominant and
significant. They could be kept under control at least in part by
inclusion of higher order corrections, in particular NNLO as well as
all-order resummation, which might be especially relevant for small-$x$ 
HERA data~\cite{Altarelli:2008aj,Caola:2010cy,Caola:2009iy}. Once
resummation corrections are properly included, it might be convenient to
use for DIS only HERA data, which are free of ambiguity related to
nuclear corrections or power-suppressed corrections. It
will be interesting to repeat the analysis presented here once PDF sets 
which include these higher order effects become available.

\vspace{0.5cm}
{\bf\noindent  Acknowledgments \\}
We thank V. Radescu and G. Watt for discussions, G.~Altarelli for a
critical reading of the manuscript, and A.~Martin for communicating to
us Ref.~\cite{Martin2011} prior to publication. We are grateful to
A.~Vicini and the 
staff of the LCM computer lab at Milan university for support and assistance 
with the theory group computing farm. M.U. is supported by the 
Bundesministerium  f\"ur Bildung and Forschung (BmBF) of the Federal 
Republic of Germany (project code 05H09PAE).
This work was partly supported by the Spanish MEC FIS2007-60350 grant.

\bigskip


\clearpage



\end{document}